\documentclass[twocolumn,amsmath,amssymb,aps,pra,superscriptaddress,showpacs,10pt,floatfix]{revtex4-2}

\usepackage{graphicx}\usepackage{grffile} \usepackage{array}

\usepackage{bbm}\usepackage{color}
\usepackage[utf8]{inputenc}
\usepackage[OT1]{fontenc}

\usepackage{units} 

\usepackage{upgreek}

\usepackage{newtxtext} 
\usepackage{newtxmath}

\usepackage{microtype}

\usepackage{dsfont}\usepackage{hyperref}\usepackage{xcolor}

\usepackage[showonlyrefs]{mathtools}		\mathtoolsset{showonlyrefs=false}

\renewcommand{\L}{\text{L}}
\newcommand{\W}{\text{W}}

\newcommand{\bra}[1]{\left\langle{#1}\right|}
 \newcommand{\ket}[1]{\left|{#1}\right\rangle}
 \newcommand{\braket}[2]{\langle{#1}|{#2}\rangle}

 \newcommand{\traceind}[2]{\operatorname{Tr}_{#1}\left\{ #2 \right\}}

\newcommand{\e}[1]{\operatorname{e}^{#1}}

\newcommand{\cn}{\text{cn}}

\newcommand{\sn}{\text{sn}}

\newcommand{\I}{i}

\newcommand{\D}{\text{d}}

\def\ba#1\ea{\begin{align}#1\end{align}}

\newcommand{\del}{\delta}
\newcommand{\taub}{\tau}

\usepackage{braket}

\usepackage{eso-pic}
\usepackage{xcolor}

\usepackage{tikz}                       \usepackage{pgfplots}                   \usetikzlibrary{spy}
\usepgfplotslibrary{external}

\usetikzlibrary{shapes,arrows,shadows,3d,calc}
\usetikzlibrary{decorations.pathmorphing,decorations.pathreplacing,decorations.markings,trees}
\usetikzlibrary{pgfplots.groupplots}

\pgfplotsset{every axis legend/.append style={cells={anchor=west},
    at={(0.96,0.04)},
    anchor=south east,
    font=\scriptsize
  },
  every axis/.append style={yticklabel style={/pgf/number format/fixed zerofill, /pgf/number format/precision=2}
  },
  width= 0.45\textwidth, height=5cm, xmajorgrids=false, xminorgrids=false, minor x tick num=1
}

\usepackage[hang]{subfigure} 

\pgfplotsset{compat=1.8}

\usepackage{booktabs}

\begin{document}
\title{High-gain quantum free-electron laser: long-time dynamics and requirements}
  \author{Peter Kling}
  \affiliation{Institute of Quantum Technologies, German Aerospace Center (DLR), S\"oflinger Stra{\ss}e 100, D-89077 Ulm, Germany}
  \affiliation{Institut für Quantenphysik and Center for Integrated Quantum Science and Technology $\left(\text{IQ}^{\text{ST}}\right)$, Universität Ulm, Albert-Einstein-Allee 11, D-89081, Germany} 
\author{Enno Giese}
\affiliation{Institut für Quantenphysik and Center for Integrated Quantum Science and Technology $\left(\text{IQ}^{\text{ST}}\right)$, Universität Ulm, Albert-Einstein-Allee 11, D-89081, Germany} 
\author{C.~Moritz Carmesin}
\affiliation{Institut für Quantenphysik and Center for Integrated Quantum Science and Technology $\left(\text{IQ}^{\text{ST}}\right)$, Universität Ulm, Albert-Einstein-Allee 11, D-89081, Germany}
\affiliation{Helmholtz-Zentrum Dresden-Rossendorf eV, D-01328 Dresden, Germany}
\author{Roland Sauerbrey}
\affiliation{Helmholtz-Zentrum Dresden-Rossendorf eV, D-01328 Dresden, Germany}
\author{Wolfgang P. Schleich}
\affiliation{Institut für Quantenphysik and Center for Integrated Quantum Science and Technology $\left(\text{IQ}^{\text{ST}}\right)$, Universität Ulm, Albert-Einstein-Allee 11, D-89081, Germany}
\affiliation{Institute of Quantum Technologies, German Aerospace Center (DLR), S\"oflinger Stra{\ss}e 100, D-89077 Ulm, Germany}
\affiliation{Hagler Institute for Advanced Study at Texas A\&M University, Texas A\&M AgriLife Research, Institute for Quantum Science and Engineering (IQSE) and Department of Physics and Astronomy, Texas A\&M University, College Station, Texas 77843-4242, USA }

\begin{abstract}
We solve the long-time dynamics of a high-gain free-electron laser in the quantum regime. 
In this regime each electron emits at most one photon on average, independently of the initial field. In contrast, the variance of the photon statistics shows a qualitatively different behavior for different initial states of the field. 
We find that  the realization of a  seeded Quantum FEL is more feasible than self-amplified spontaneous emission.     
\end{abstract}

\maketitle

\section{Introduction}
\label{sec:Introduction}

In the quantum regime of the free-electron laser (FEL) the electrons undergo discrete scattering events instead of following continuous trajectories~\cite{carmesin20}. More precisely, an electron in the Quantum FEL~\cite{schroeder,boni06,pio_prl,boni17,serbeto09,brown17, anisimov18,gaiba} occupies only \textit{two} discrete momentum levels~\cite{NJP2015} in analogy to an atomic laser~\cite{scullylamb}. 

By identifying three constants of motion we solve in the present article the long-time dynamics of a Quantum FEL in the high-gain regime, within (i) an analytical approximation and (ii) a numerical simulation. Moreover, we discuss fundamental requirements~\cite{debus} to realize such a device in an experiment.       

Employing momentum-jump operators  we showed in Ref.~\cite{PRA2019} that the dynamics of a high-gain Quantum FEL is effectively governed by the Dicke Hamiltonian~\cite{dicke}. We briefly review this model in Sec.~\ref{sec:Analogy_to_angular_momentum}. 
In order to solve the resulting equations of motion we applied in Ref.~\citep{PRA2019} a parametric approximation~\cite{siegman} and obtained an exponential growth of the laser intensity along the wiggler length.    
However, this approximation breaks down when the number of emitted photons becomes large after a certain interaction time. Therefore, we derive in Sec.~\ref{sec:Long_time_solution}  solutions beyond the short-time limit.

As a first result of these studies, we find that in the quantum regime each electron emits at most only a single photon~\cite{boni06}, 
in contrast to the multi-photon processes dominating the classical regime~\cite{mciver79quantum}. Moreover, we
derive expressions for the saturation length and consider deviations from resonance. 

Secondly, we discuss the variance of the photon number. For a start-up from vacuum, the field possesses almost chaotic statistics. In case of a seeded Quantum FEL, the behavior of the variance depends strongly  on the initial field state. A Fock state or a coherent state with a high photon number lead to comparably narrow photon distributions in the course of time. In contrast, the statistics evolving from a thermal state remains broad, but becomes much narrower than a thermal distribution when the intensity assumes its maximal value.

Our results allow us in Sec.~\ref{sec:Experimental_Requirements} to identify the challenges for a Quantum FEL experiment, and to explain the necessity for an optical undulator~\cite{boni05,steiniger}. Indeed, it was argued in Ref.~\cite{debus} that the combined influence of space charge and spontaneous emission into all modes prevents an effective Quantum FEL operation for more than several gain lengths which drastically reduces the possible laser intensity. We can circumvent this loss of intensity if we consider a seeded FEL instead of self-amplified spontaneous emission (SASE). This effect is a direct consequence of the decreased saturation length in a seeded Quantum FEL so that one drops below the problematic limit put forward in Ref.~\cite{debus}.

Finally, we summarize our main results and conclude in Sec.~\ref{sec:Conclusions}.
To keep this article self-contained we add the Appendices~\ref{sec:Analytic_approximation} and~\ref{sec:Numerical_solution} which are devoted to the detailed calculations associated with our analytical approximation and the numerical approach, respectively.

 \section{Quantum FEL: Basic Building Blocks}
  \label{sec:Analogy_to_angular_momentum}

In Ref.~\citep{PRA2019} we have formulated the FEL dynamics~\cite{becker83,fernandez87} in terms of collective jump operators for the electron momentum. In the following, we review this description, where each electron populates levels on a discrete momentum ladder induced by the quantum mechanical recoil $q\equiv2\hbar k $ the electron experiences when it scatters from the fields. 
Here $\hbar$ denotes the reduced Planck constant while $k$ is the wave number of the laser  and wiggler field in the co-moving Bambini-Renieri frame~\cite{bambi,*brs} where both coincide.

\subsection{Collective Hamiltonian}

By performing a rotating wave-like approximation \citep{PRA2019} we found that the motion of each electron in the quantum regime is restricted to only two momentum levels, 
that is the excited state with momentum $p$, close to the resonant momentum $p=q/2$, and the ground state with $p-q\cong -q/2$. In the language of collective operators we therefore defined
\begin{equation}\label{eq:J_pm} 
 \sum\limits_{j=1}^N\ket{p-q}^{(j)}\bra{p}\equiv \hat{J}_- \equiv \left(\hat{J}_+\right)^\dagger
\end{equation}
and
\begin{equation}\label{eq:J_z}
      \frac{1}{2}\sum\limits_{j=1}^N\left(\ket{p}^{(j)}\bra{p}-\ket{p-q}^{(j)}\bra{p-q}\right)
       \equiv  
       \hat{J}_z\,,
\end{equation}
where we sum over projection operators for  all $N$ electrons with $\ket{p}^{(j)}$ denoting the momentum eigenstate of the $j$-th electron.

Indeed, $\hat{J}_+$, $\hat{J}_-$, and $\hat{J}_z$  satisfy the commutation relations for angular momentum~\cite{boniprep69,*boniprep70,ct}, that is $[\hat{J}_+,\hat{J}_-]=2\hat{J}_z$ and 
$[\hat{J}_z,\hat{J}_\pm]=\pm\hat{J}_\pm$.
Hence, we identify the jump operators with the  ladder operators $\hat{J}_{+}$ and $\hat{J}_{-}$ of angular momentum, while $\hat{J}_z$ describes its $z$-component~\footnote{We emphasize that these operators agree only formally with angular momentum, but do not describe the  angular momentum of the electrons.}.

The restriction to two momentum levels leads us to the dimensionless Dicke Hamiltonian~\cite{dicke}
\begin{align}\label{eq:Hdicke_ang}
\hat{H}_\text{eff}\equiv\varepsilon\left(\hat{a}_\L \hat{J}_+ +\hat{a}_\L^\dagger \hat{J}_-\right)-\del \, \hat{n}\,,
\end{align}
where the dynamics of the laser mode is described by the photon annihilation, creation, and number operator, $\hat{a}_\L$, $\hat{a}_\L^\dagger$, and $\hat{n}\equiv \hat{a}_\L^\dagger \hat{a}_\L$, respectively.  
These field operators fulfill the commutation relation $[\hat{a}_\L,\hat{a}_\L^\dagger]=1$.

Moreover, we have recalled the dimensionless parameter $\varepsilon\equiv g/\omega_\text{r}$ as the ratio of the coupling constant $g$ for electrons and fields~\cite{becker82,NJP2015} and the recoil frequency~\cite{NJP2015}
$\omega_\text{r}\equiv q^2/(2m\hbar)$ with $m$ denoting the mass of an electron. 
In addition, $\del \equiv (p-q/2)/(q/2)$
describes the scaled detuning of the initial momentum $p$ of the electrons from resonance at 
$p=q/2$. We emphasize that also the time variable $\tau \equiv \omega_\text{r}t$  in this description is in a dimensionless form.

The approximation leading to the effective Hamiltonian of Eq.~\eqref{eq:Hdicke_ang} is only valid, provided the quantum parameter $\alpha_N$ (compare to Table~\ref{tab:bambi_lab}) satisfies the inequality
\begin{align}\label{eq:alpha_def} 
\alpha_N\equiv \frac{g\sqrt{N}}{\omega_\text{r}}\ll 1\,,
\end{align}
which defines the quantum regime of the FEL. 

Moreover, we require that the detuning from resonance is small, that is $\del <1$. For a realistic electron beam, where each electron has a different initial momentum we additionally require $\Delta p <q$ for the initial momentum spread $\Delta p$ of the electrons.

\subsection{Constants of motion}

Despite of several attempts~\cite{tavcumm,boniprep69,*boniprep70,scharf74,kumar,gambini,caruso,les_houches_82} no closed analytic solution for the dynamics dictated by the Dicke Hamiltonian has been found. Hence, we  resort to approximations and numerical methods. Hereby, we closely follow the lines of the existing literature~\cite{kumar,boniprep69,*boniprep70,gambini,walls70} on  this problem. In addition, we study the effect of a nonzero detuning from resonance. 

The key ingredient of our approach is the identification of three constants of motion~\cite{kumar}.
From the Hamiltonian in Eq.~\eqref{eq:Hdicke_ang} we straightforwardly  derive  that the total angular momentum 
\begin{equation}\label{eq:qhigh_AB} 
\hat{A} \equiv\hat{\boldsymbol{J}}^2=\frac{1}{2}\left(\hat{J}_{+}\hat{J}_{-}+\hat{J}_{-}\hat{J}_{+}\right) +\hat{J}_z^2
\end{equation}
as well as the total number of excitations
\begin{equation}
\hat{B} \equiv \hat{n} +\hat{J}_z
\end{equation}
are constants of motion which commute with each other, that is $\left[\hat{A},\hat{B}\right]=0$. 

Moreover, the Hamiltonian  
\begin{equation}
\hat{C}\equiv\hat{H}_\text{eff}
\end{equation}
is independent of time and thus itself constitutes a third constant. 

\subsection{Our Approaches}
We use these quantities to investigate the long-time dynamics of a high-gain Quantum FEL in an analytical as well as a numerical approach. The former one is carried out in detail in App.~\ref{sec:Analytic_approximation} and relies~\cite{kumar} on the decoupling of the Heisenberg equations of motion of the photon-number operator $\hat{n}$ with the help of $\hat{A}$, $\hat{B}$, and $\hat{C}$. To solve the resulting differential equation of non-commuting operators we approximate the operators by c-numbers~\cite{kumar,gambini}. By this procedure we find an analytical expression for the mean photon number $n\equiv\braket{\hat{n}}$ in terms of a Jacobi elliptic function~\cite{gambini,boniprep70,byrd}. 

In contrast, the approach of App.~\ref{sec:Numerical_solution} towards a numerical solution~\cite{walls70} is  based on the evolution of time-dependent state vectors as solutions of the 
Schr\"odinger equation. For that, we expand the state of the laser field into Fock states $\ket{n}$ with photon number $n$ while the collective state of the electrons is described by the basis $\ket{r,m}$, where $r$ corresponds to the total angular momentum and $m$ to its $z$-component. Since $\hat{A}$ and $\hat{B}$ are constant, the set $n,r,m$ of three independent quantum numbers is reduced to a single one. As a consequence, the Schr\"odinger equation leads to a three-term recurrence relation for the expansion coefficients which we straightforwardly solve by the  numerical diagonalization of a tri-diagnoal matrix of dimension $(N+1)\times(N+1)$.

 \section{Long-time dynamics}
\label{sec:Long_time_solution}

In the following we present our results of  the analytical approximation and the numerical simulation for the long-time dynamics of a high-gain Quantum FEL. Here we first focus on the time evolution of the mean number of photons and concentrate on the dependence of the saturation intensity and length on the number of electrons. We conclude by discussing the variance of the photon distribution of the Quantum FEL.

\subsection{Time evolution of mean photon number}

In App.~\ref{sec:Analytic_approximation} we derive the approximate expression
\begin{equation}\label{eq:nL_jacob} 
n(L)=n_0+(n_+-n_0)\,\cn^2\left[\frac{L}{2L_g}\sqrt{\frac{n_++n_-}{N}}-K,\mathfrak{K}\right]
\end{equation}
for the mean photon number $n\equiv\braket{\hat{n}}$ of a Quantum FEL as a function of the length $L$ of the wiggler, where $\cn$ denotes a Jacobi elliptic function~\cite{byrd}. The quantities $n_+$  and $n_-$ are roots of the right-hand side of the differential equation~\eqref{eq:qhigh_dn2} for $n$. The explicit expressions for these roots are given in Eq.~\eqref{eq:qhigh_npm} and depend on the parameters of this differential equation, that is the initial photon number $n_0$, the electron number $N$, and the detuning $\del/\alpha_N$ from resonance in units of the quantum parameter.  

Moreover, we have recalled from Refs.~\cite{PRA2019,boni06} the gain length
\ba\label{eq:L_g_def} 
L_g\equiv\frac{c}{2g\sqrt{N}}
\ea
of a Quantum FEL. 

In addition, the Jacobi elliptic function $\text{cn}$ is characterized by the modulus 
$\mathfrak{K}=\mathfrak{K}(n_+,n_-)$ from Eq.~\eqref{eq:app_jac_ksubst}, whereas $K\equiv K(\mathfrak{K})$ describes a complete elliptic integral of first kind~\cite{byrd}.

\begin{figure}
\includegraphics[scale=1]{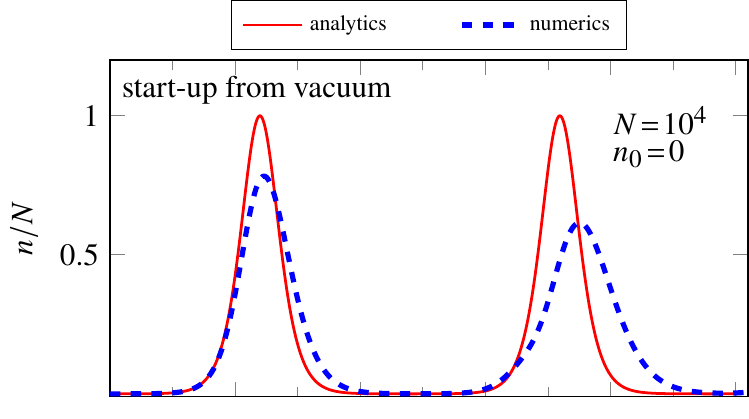}
\includegraphics[scale=1]{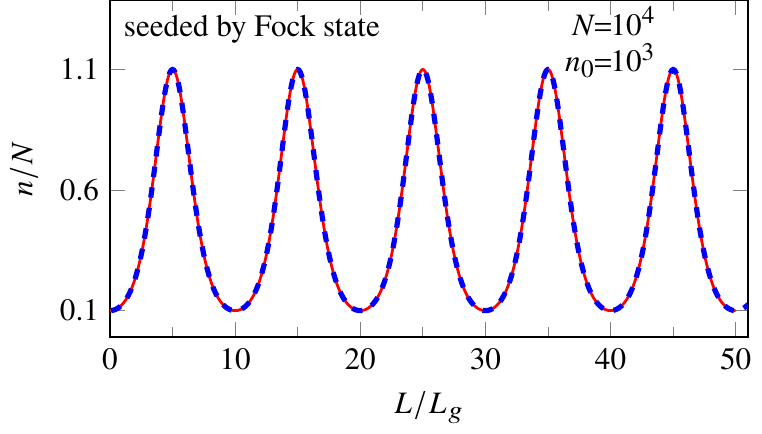}
\caption{Mean photon number $n=\braket{\hat{n}}$ of a high-gain Quantum FEL as a function of the wiggler length $L$ scaled in units of the gain length
$L_g$: 
We compare the analytical approximation (red line) from Eq.~\eqref{eq:nL_jacob} to the numerical solution (blue, dashed line) of Eq.~\eqref{eq:qhigh_three_term} for the start-up from vacuum (top), 
that is $n_0=0$, as well as for a seeded FEL (bottom) with $n_0=10^3$ photons in an initial Fock state. For both plots we have assumed resonance, $p=q/2$, and the value $N=10^4 $ for the number of electrons. In all cases the curves show an exponential growth for short times which saturates and leads to a first maximum followed by further oscillations.
For the start-up from vacuum the photon number saturates at about $10$ gain lengths. There,  the approximate solution takes on the value $n=N$, that is each electron has emitted exactly one photon, while the numerically computed maximum  is at about $n \cong 0.8 N$. In addition, the locations of these maxima for analytics and numerics are slightly shifted with respect to each other. For an FEL seeded by a large Fock state, numerics and analytics agree very well. In this case, we obtain complete oscillations of the mean photon number between the values $n_0$ and $n_0+N$ with a significantly shorter periodicity than for the start-up from vacuum.}
\label{fig:numvsanal}
\end{figure}

We emphasize that the initial state of the laser field enters in the approximation, Eq.~\eqref{eq:nL_jacob}, only through the mean photon number $n_0$. In Fig.~\ref{fig:numvsanal} we compare the analytical approximation for the mean photon number $n=\braket{\hat{n}}$ to the numerical solution, for exact resonance $p=q/2$ and with $N=10^4$ 
electrons~\footnote{ Based on the  formalism of second quantization of G.~Preparata,\ Quantum field theory of  the free-electron laser,
 Phys. Rev. A\ \textbf {38},\ 233\ (1988), analogous results for the mean photon number were shown in Ref.~\cite{gaiba}, at least for the start-up from vacuum.  
In addition, we study in our article a seeded Quantum FEL and consider higher moments of the photon statistics as well as a nonzero detuning $\del$ from resonance.}.
Here we consider the start-up from vacuum (top) as well as a seeded FEL evolving from a Fock state with $n_0=10^3$ photons (bottom).

In both cases we observe an exponential start-up in accordance with Ref.~\cite{PRA2019}. However, this growth saturates for increasing values of the wiggler length $L$ leading to a local maximum of the photon number. The mean photon number then decreases until it reaches its initial value, before it again increases in an oscillatory-like behavior.       

For start-up from vacuum we deduce from the analytical approach that each electron emits at most one photon, that is $n_\text{max}=N$, in contrast to  the smaller  value $n_\text{max}\cong 0.8\,N$ obtained by the numerical simulation. 
The second maximum of the numerical solution, however, is even stronger suppressed compared to the analytical approximation.  

We interpret these deviations as the result of entangled Dicke states for the electron momenta~\citep{dicke,PRA2019} which the numerical solution takes into account. The  oscillations of the analytical solution between $0$ and $N$ indicate that in this model all electrons are in the ground state  when the maximum photon number is reached. In the exact treatment, however, the electrons entangle with each other due to their common interaction with the laser field. This entanglement prevents a product state, where each electron is in the ground state, decreasing the maximum photon number.

Despite these differences between the numerical and the approximate results,
the latter one  is helpful  to estimate the qualitative behavior of the dynamics. This feature becomes even more important in the discussion of experimentally relevant length scales in Sec.~\ref{sec:Experimental_Requirements}, when we increase the electron number $N$ to more realistic values where a numerical diagonalization becomes impracticable. In this limit, we have to resort to the predictions given by the expression in Eq.~\eqref{eq:nL_jacob}.

In contrast, for a seeded Quantum FEL the numerical and analytical solution for the mean photon number  agree very well, at least for a Fock state. In this situation we observe oscillations of the mean photon number between $n_0$ and $n_0+N$.
We note that the periodicity of these oscillations is much shorter than for the start-up from vacuum.

\begin{figure}
\centering
\includegraphics[scale=1]{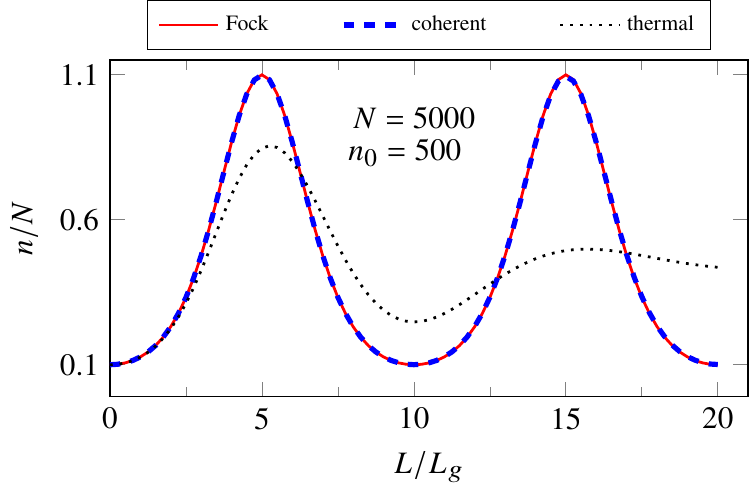}

\caption{Mean photon number $n=\braket{\hat{n}}$ of a seeded high-gain Quantum FEL as a function of the wiggler length $L$ in units of the gain length
$L_g$: We depict the  numerical solution for $n$ with respect to three different initial states for the laser field, that is a Fock state (red line), a coherent state (blue, dashed line), and a thermal state (black, dotted line). The latter two states are described by the photon distributions in Eqs.~\eqref{eq:pn_coh} and~\eqref{eq:pn_therm}. In all three examples we have chosen the values $n_0=500$ and $N=5000$ for the initial mean photon number and the number of electrons, respectively. While a Fock state and a coherent state lead to the same oscillatory behavior of the photon number between $n_0$ and $N+n_0$, the situation for a thermal state is different: Here the first maximum of $n$ attains the decreased value of approximately $0.8 N$ instead of $1.1 N$. Moreover, we do not observe  full oscillations, but the photon number seems to approach a constant value of approximately $0.5 N$.}
\label{fig:mean_seeded}
\end{figure}

So far, we have only considered the case of the initial state of the laser field being given by a Fock state.
We now study the influence 
of the photon statistics  of the initial state on the time evolution of the mean photon number and refer to App.~\ref{sec:Numerical_solution} for details.

Therefore, we show in Fig.~\ref{fig:mean_seeded} the numerically evaluated expectation value $n=\braket{\hat{n}}$ as a function of $L$, with the initial state of the field given by (i) a Fock state, (ii) a coherent state, and (iii) a thermal state. To make a meaningful comparison we have chosen in the three cases the same initial mean photon number $n_0=500$ while assuming $N=5000$ electrons. A coherent state is usually employed to model the output of a coherent light source such as a laser. In contrast, a thermal state describes a random, incoherent source, for example a light bulb~\cite{schleich,meystre}.

We observe that a coherent state leads approximately to the same behavior as a Fock state, that is an oscillation of the photon number between $n_0$ and $N+n_0$. However, the curve corresponding to a thermal input state is different: Here the maximum photon number reaches only the  decreased value of $0.8 N$ instead of $1.1 N$. In addition, the oscillatory behavior is washed out and $n$ seems to approach a constant value of about $0.5 N$.     

Similar to a Fock state, a coherent state with a high photon  number possesses a sharply peaked photon distribution $p_n$. We obtain the mean photon number through
\ba
n(L)=\sum\limits_{n'=0}^\infty p_{n'}\, n_{n'}(L)
\ea 
expressed by Eq.~\eqref{eq:fn_av}. Here $n_{n'}=n_{n'}(L)$ denotes the mean photon number for the initial Fock state $n'$. 

Hence, mainly contributions with photon numbers close to the mean $n_0$ are relevant. Since the dynamics of these contributions occurs on similar time scales, the averaging process yields results identical to the ones originating from a delta-peaked photon distribution, that is, for a Fock state.  

In contrast, the photon distribution of a thermal state is very broad and, moreover, is not symmetric around $n_0$~\cite{meystre}. Hence, we have to average over many differently weighted contributions
which corresponds to many different time scales of the resulting dynamics.

Additionally, the probabilities $p_n$ for a thermal state increase for photon numbers close to zero. 
There, the dynamics of $n$ is drastically different from high initial photon numbers as apparent in Fig.~\ref{fig:numvsanal}. This mixing of many different time scales, combined with the influence of the qualitatively different behavior for small photon numbers explains the difference in Fig.~\ref{fig:mean_seeded} between a thermal input state and  a coherent or a Fock state.

\subsection{Saturation intensity and length }

In this section we investigate the magnitude $n_\text{max}$ and the position $L_\text{max}$ of the first maximum of the mean photon number shown in Fig.~\ref{fig:numvsanal}, that is the saturation intensity and the saturation length, respectively, of a Quantum FEL. Our analysis distinguishes the two cases of exact resonance and near resonance.

\subsubsection{Resonant case}

For the time being, we consider exact resonance, that is $p=q/2$, and assume that the laser field is initially described by a Fock state. 
From Eqs.~\eqref{eq:nL_jacob} and~\eqref{eq:qhigh_npm} we derive the simple relation $n_\text{max}\cong n_0+N$ for the maximum photon number which we have already deduced from Fig.~\ref{fig:numvsanal}. 
There, we have also found that for $n_0=0$ the numerical result is lower than the estimated one, at least for our choice of $N=10^4$ electrons. 

\begin{figure}

\includegraphics[scale=1]{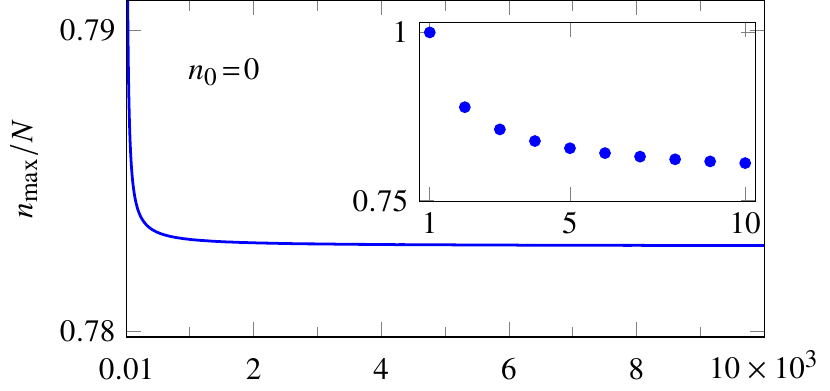}

\includegraphics[scale=1]{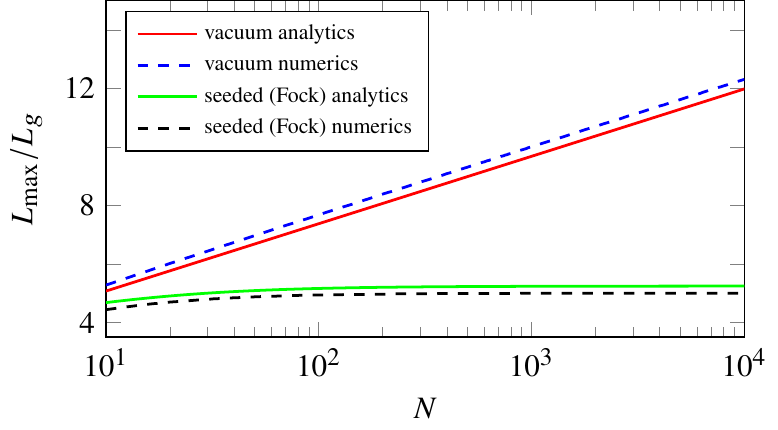}

\caption{The first maximum $n_\text{max}$ (top) of the photon number and the corresponding saturation length $L_\text{max}$ (bottom), both as functions of the number $N$ of electrons  and  for resonance $p=q/2$. 
In the top panel, which covers only the start-up from vacuum with $n_0=0$, we observe that $n_\text{max}$  very quickly reaches the constant value of $n_\text{max}=0.78\,N$ 
in contrast to  the analytical approximation which predicts $n_\text{max}=N$.
The inset shows the behavior of $n_\text{max}$ for very small values of $N$.
For $N=1$ we find that the electron emits one photon due to the Rabi oscillation in the single-electron approach of Ref.~\cite{NJP2015}. At the bottom, the comparison of analytics, Eq.~\eqref{eq:length_seeded}, (red line) to numerics (blue, dashed line) reveals  the same logarithmic behavior of $L_\text{max}$ apart from a small shift between the two curves, if we consider start-up from vacuum, $n_0=0$. In the case of a Quantum FEL seeded with a Fock state and with a fixed ratio of $n_0/N=0.1$  analytics (green line) and numerics (black, dashed line) also agree well. Here both curves  predict the constant value $L_\text{max}\cong 5 L_g$ with $L_g$ denoting the gain length from Eq.~\eqref{eq:L_g_def}.}
\label{fig:Lmax_N_num}
\end{figure}

In the upper part of Fig.~\ref{fig:Lmax_N_num} we now examine if this discrepancy between analytics and numerics continues to exist  for different  values of $N$ by plotting the numerically evaluated result for $n_\text{max}$ against $N$. In the limit of a single electron exactly one photon is emitted in accordance with the single-electron approach of Ref.~\cite{NJP2015}. Here Rabi oscillations between excited and ground state occur and no entangled Dicke states, which we have identified as the reason for a decreased ratio $n_\text{max}/N$, emerge. 

However, already for low electron numbers the value $n_\text{max}\cong 0.78N$ is reached which then remains constant for the whole range of the considered values for $N$. We therefore deduce that (i) this behavior continues for increasing values of $N$ where numerics starts to become impracticable, and that (ii) the analytical result $n_\text{max}=N$ provides us at least with the correct order of magnitude. Already in Ref.~\cite{scharf74} it was predicted that the maximum photon number $n_\text{max}$ resulting from the Dicke Hamiltonian with $N\gg 1$ atoms varies  between $n_\text{max} \cong n_0+N$ for a large initial photon number $n_0\gg 1$ and $n_\text{max}\cong 4/5$ for $n_0=0$ in accordance with our numerical solution.       

In Fig.~\ref{fig:numvsanal} we have also observed a different $L_\text{max}$ for the start-up from vacuum compared to the seeded case. To find an analytical expression for $L_\text{max}$ 
we now take a closer look at the elliptic  function $\cn$  in Eq.~\eqref{eq:nL_jacob}.
The first maximum of this function occurs, when its arguments vanish. With the help of this condition and the asymptotic behavior~\cite{byrd} of $K$ in Eq.~\eqref{eq:app_K_asympt}
for $\mathfrak{K}=1-\mathcal{O}(N^{-1})\rightarrow 1$ we arrive at the expression
\ba\label{eq:length_seeded} 
\frac{L_\text{max}}{L_g}\cong 2\ln{\left(4\sqrt{\frac{N}{n_0+1}}\right)} 
\ea
for the saturation length with $p=q/2$ and $N\gg 1$~\footnote{In the classical regime the ratio of saturation and  gain length is typically independent of the electron number, at least for SASE with $L_\text{max}^{(\text{cl})}\cong 4\pi\sqrt{3}L_g^{(\text{cl})}$~\cite{schmueser}. However, the gain and saturation mechanisms in the quantum regime differ from the classical ones. While for a Quantum FEL population inversion of the excited and ground state is crucial, the dynamics of its classical counterpart is based~\cite{schmueser} on the micro-bunching of the electron beam.}.\nocite{schmueser}

For $n_0=0$ we thus read off a logarithmic growth of $L_\text{max}$ with the electron number $N$, that is $L_\text{max}\propto \ln{N}$. 
In the case of a seeded Quantum FEL with $n_0 \gg 1$ the important parameter that determines the magnitude of $L_\text{max}$ is the \textit{ratio} of electron number $N$  and initial photon number $n_0$. 
This scaling explains the difference in Fig.~\ref{fig:numvsanal} of $L_\text{max}$ for a seeded FEL,  and one starting from vacuum  because of $N/n_0 \ll N$. 
 
In the lower part of Fig.~\ref{fig:Lmax_N_num} we depict $L_\text{max}$ as a function of $N$. Here we compare the analytical expression in Eq.~\eqref{eq:length_seeded} to the numerical simulation. 
For the start-up from  vacuum  we observe that both curves show the same behavior, apart from a small shift, already apparent in Fig.~\ref{fig:numvsanal}. 
Hence, we infer that the analytical approximation gives a reasonable estimate also for $L_\text{max}$.   
For a seeded FEL with the fixed ratio $n_0/N=0.1$, the numerical and the analytical curve also match very well with both predicting a constant value of $L_\text{max}$, in accordance with Eq.~\eqref{eq:length_seeded}.

\subsubsection{Off-resonant case}

\begin{figure}
\centering
\includegraphics[scale=1]{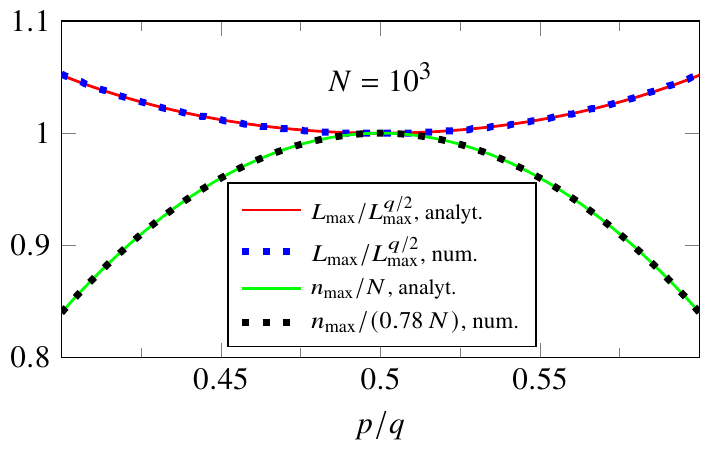}

\caption{Saturation length $L_\text{max}$ (upper curves) and the corresponding maximum photon number $n_\text{max}$ (lower curves), as functions of the momenta $p$ in the neighborhood of resonance for $N=10^3$ electrons and for $n_0=0$. 
We observe that $n_\text{max}$ \textit{decreases} and $L_\text{max}$ \textit{increases} for a growing deviation from resonance $p=q/2$. 
For both, $L_\text{max}$ and $n_\text{max}$, we find perfect agreement between analytics, that is the red curve and the green curve, respectively, and numerics, that is the blue dotted and the black dotted curve, respectively. 
We emphasize that the analytical and the numerical curves are normalized to their respective values at resonance which, however, differ from each other. 
For example, the analytical result for $n_\text{max}$ is divided by $N$ while its numerical counterpart is divided by $0.78\,N$.  }
\label{fig:kappa_max_num}
\end{figure}

We now study the dependence of the maximum photon number on the detuning of the momentum of the electrons from resonance. For that, we restrict ourselves to the start-up from vacuum $n_0=0$. 

According to the analytical solution from Eq.~\eqref{eq:nL_jacob}, the maximum photon number $n_\text{max}$  is given by the elementary relation 
\ba\label{eq:qhigh_nmax}
n_\text{max}=n_+\cong N\left(1-\frac{\del^2}{4\alpha_N^2}\right)
\ea
valid for $N\gg1$, where we have used the expression for $n_+$ from Eq.~\eqref{eq:qhigh_npm}. 

Thus, $n_\text{max}$ is different from zero only for $-2\alpha_N<\del < 2\alpha_N$, which gives rise to a gain bandwidth of $q/(2\alpha_N)$ in momentum space, in accordance with the exponential-gain regime~\cite{PRA2019}. 

Moreover, from Eq.~\eqref{eq:app_K_asympt} we derive for $N\gg 1$ the asymptotic expression
\begin{equation}\label{eq:qhigh_Lmax} 
\frac{L_\text{max}}{L_\text{g}}=\frac{1}{ \sqrt{1-\frac{\del^2}{4\alpha_N^2}}}
\left[\ln{N}+4\ln{2}+2\ln{\left(1-\frac{\del^2}{4\alpha_N^2}\right)}\right]
\end{equation}
for the saturation length.

In Fig.~\ref{fig:kappa_max_num} we show both $n_\text{max}$ and $L_\text{max}$ as a function of $p$ which reveals a perfect agreement between our analytical results and numerics. 
We note that $n_\text{max}$ \textit{decreases} for an increasing deviation from resonance $p=q/2$, while the value of $L_\text{max}$  \textit{increases}.

\subsection{Variance of photon number}

A quantum mechanical observable is not only characterized by its mean value, but also by its higher moments. 
The numerical solution of Eq.~\eqref{eq:qhigh_three_term} enables us to calculate these moments for the observables of a high-gain Quantum FEL. 
We now study the second moment, that is the variance of the photon number which is a measure of the intensity fluctuations of the emitted radiation.

\begin{figure}
\centering
\includegraphics[scale=1]{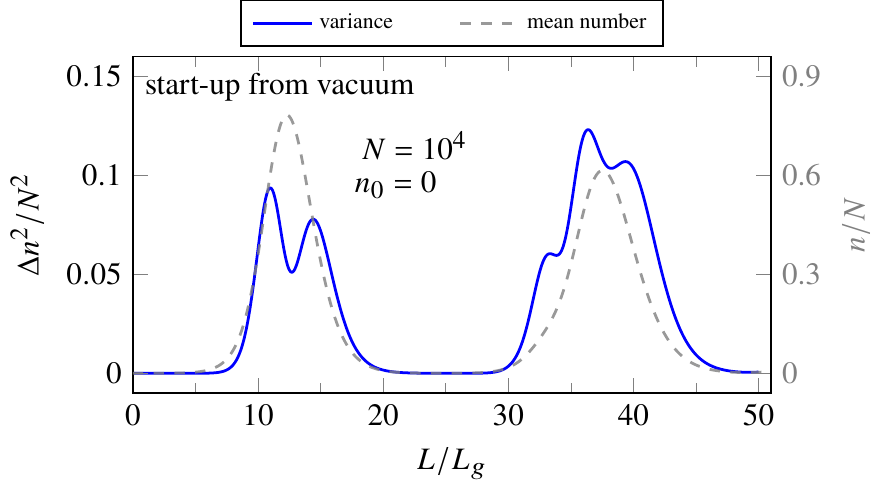}

\caption{Variance $\Delta n^2$ of the photon-number distribution in a high-gain Quantum FEL as a function of the wiggler length $L$ scaled in units of the gain length $L_g$. For the underlying numerical simulation we have assumed resonance $p=q/2$, $N=10^4$ electrons, and start-up from vacuum, that is $n_0=0$.  We obtain a qualitatively similar behavior as for the mean photon number (compare to gray, dashed line and right axis), that is, exponential growth, local maximum, and decrease in an oscillatory-like fashion. However, the structure of $\Delta n^2$ is more complicated. For example, close to  $L_\text{max}$ a dip occurs, where  $\Delta n^2\cong0.05\, N^2$ while $n_\text{max}\cong0.8\,N$. Hence, the value of $\Delta n^2$ is smaller, but roughly of the order of magnitude of $n_\text{max}^2$ corresponding to an almost chaotic behavior of the laser field.}
\label{fig:variance_high_long}
\end{figure}

In Fig.~\ref{fig:variance_high_long} we depict the variance 
\ba 
\Delta n^2\equiv \braket{\hat{n}^2}-\braket{\hat{n}}^2
\ea
of the photon number as a function of the wiggler length $L$ for resonance $p=q/2$, and for the start-up from vacuum with $N=10^4$ electrons. 
Similar to the mean value  in Fig.~\ref{fig:numvsanal}, the variance shows an oscillatory behavior~\cite{scharf74}. However, compared to the mean value (dotted line) the curve corresponding to the variance displays a richer structure with a dip close to $L_\text{max}$ .
Here we find $\Delta n^2\cong 0.05 N^2$ while $n_\text{max}\cong 0.8 N$. 
Hence, the variance is smaller, but of the same order of magnitude, as the square of the mean value and we deduce a nearly chaotic behavior of the radiation field~\cite{siegman}. 

\begin{figure}
\centering
\includegraphics[scale=1]{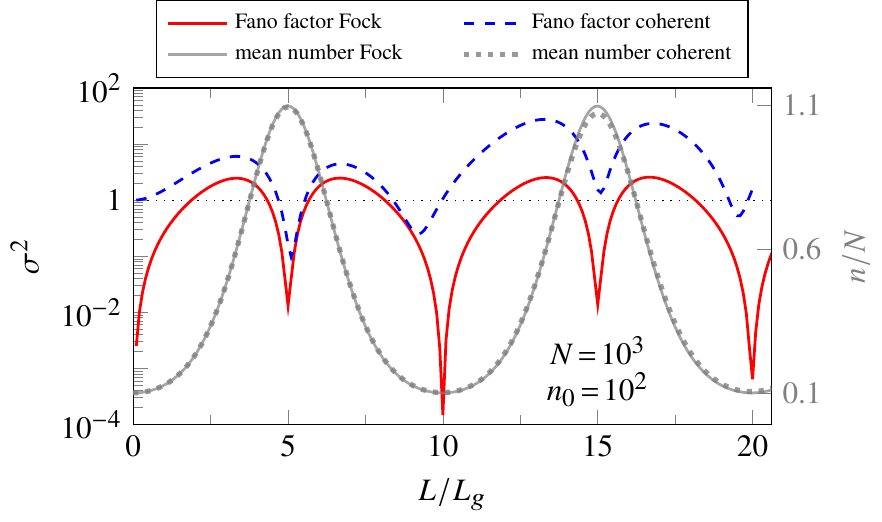}

\caption{Fano Factor $\sigma^2\equiv\Delta n^2/\braket{\hat{n}}$ of the photon-number distribution in a seeded high-gain Quantum FEL as a function of the wiggler length $L$ scaled in units of the gain length $L_g$ for two different initial states, that is a Fock state (red line)  and a coherent state (blue, dashed line). For the underlying numerical simulations we have assumed resonance $p=q/2$, $N=10^3$ electrons, and an initial mean photon number of $n_0=10^2$.
In both cases we observe super- as well as sub-Poissonian behavior which are separated by the horizontal dotted line at $\sigma^2=1$. We note that the first \textit{minimum} of this normalized variance occurs in the vicinity of the first \textit{maximum} of the mean photon number (compare to the gray solid and dotted lines as well as to the right axis) at $L\cong 5 L_g$. Moreover, we find that the fluctuations for the initially coherent case drastically increase at around $10$ gain lengths, while only slowly increasing for the case of a Fock state.}
\label{fig:var_coh_fock}
\end{figure}

The situation, however, changes, when we consider a seeded FEL illustrated in Fig.~\ref{fig:var_coh_fock}.
Here we compare the Fano Factor~\cite{schleich} $\sigma\equiv\Delta n^2/\braket{\hat{n}}$ depending on the wiggler length $L$  for two different initial states of the radiation field, 
that is a Fock state~\cite{scharf74}  and a coherent state characterized by the same mean number of photons.

While the variance for a Fock state vanishes, a coherent state possesses a Poissonian photon statistics with $\Delta n^2=\braket{\hat{n}}$. As time evolves we obtain in both cases super- but also sub-Poissonian photon statistics, in contrast to the broad distribution originating from the start-up from vacuum displayed in Fig.~\ref{fig:variance_high_long}. 
Both curves show a sub-Poissonian statistics close to the first maximum of the mean photon number indicating a non-classical state of light.
Moreover, we obtain a drastic increase of the fluctuations for larger values of $L$, when the field initially was in a coherent state. In contrast, for an initial Fock state a growth of the fluctuations is hardly visible.  

\begin{figure}
\centering
\includegraphics[scale=1]{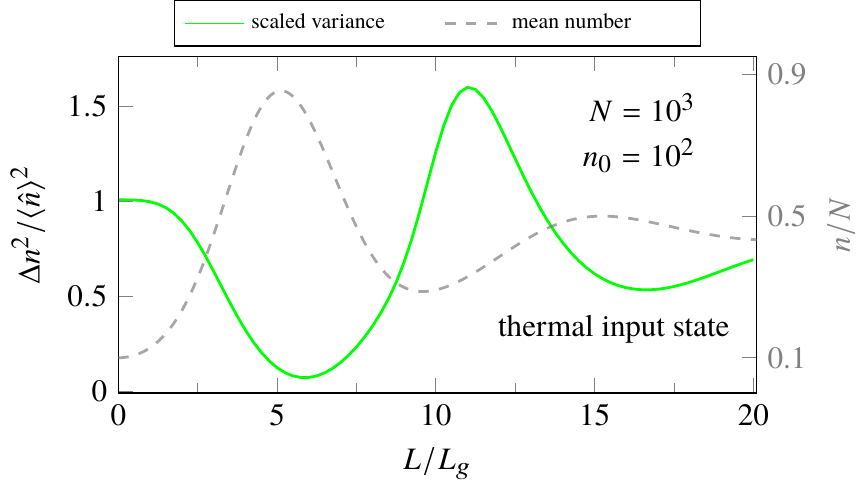}

\caption{Variance $\Delta n^2$ of the photon number in a seeded high-gain Quantum FEL as a function of the wiggler length $L$ scaled in units of the gain length $L_g$. We have normalized the variance with respect to $\braket{\hat{n}}^2$ to investigate the evolution of the fluctuations in an initial thermal state. The variance $\Delta n^2$ is of  the same order of magnitude as $\braket{\hat{n}}^2$. However, this ratio attains a significant minimum when the first maximum of the mean photon number is reached (compare to gray, dashed line and right axis).}
\label{fig:var_therm}
\end{figure}

In Fig.~\ref{fig:var_therm} we study the second moment of the field when it starts from a thermal state where $\Delta n^2 =\braket{\hat{n}}^2$. 
Although the variance of the  photon number remains of the same order of magnitude as the square of the mean value, 
the corresponding ratio attains a prominent minimum close to the saturation length where the mean intensity assumes a maximum. 
Here the corresponding statistics with $\Delta n^2 < \braket{\hat{n}}^2$ deviates significantly from the thermal distribution of the initial state. Hence, the intensity noise of the emitted light is at least partially decreased.

 \section{Experimental Requirements}
\label{sec:Experimental_Requirements}

In this section we address the challenges associated with realizing a Quantum FEL.
Already from our elementary one-dimensional description we can establish the most relevant experimental conditions.  
Constraints which go beyond the limits of our  model are only mentioned here, but are studied in more detail in Ref.~\cite{debus}.      
 
\begin{table*}
\caption[newline]{
Parameters of a Quantum FEL in the Bambini--Renieri (BR) frame (left), in the universal scaling of Ref.~\cite{boni06} (center), and in the laboratory frame (right).
We consider the five adjustable parameters wavelength $\lambda_\W$ of the undulator, its dimensionless field amplitude $a_0$, electron density $n_\text{e}$, dimensionless energy $\gamma_0$ of the electrons, and associated spread $\Delta \gamma_0$.
Here, $\lambda_\text{C}$ and $r_\text{e}$ are the Compton wavelength of the electron and the classical electron radius, respectively.
In addition, we have introduced the longitudinal dimensionless energy $\gamma$ of the electrons in the wiggler which is connected by the relation $\gamma=\gamma_0 (1+a_0^2)^{-1/2}$ to the free energy and the wiggler parameter.
}
\centering
\begin{tabular}{@{}lllcr@{}}\toprule
 & & BR  frame& universal scaling & laboratory frame \\ \midrule
quantum parameter & $\alpha_N$  & $\displaystyle \frac{g\sqrt{N}}{\omega_\text{r}} $ & $\bar{\rho}^{3/2}$ & $\displaystyle \frac{\sqrt{r_\text{e} n_\text{e}}}{32 \gamma_0^3 \sqrt{\pi}}\frac{\lambda_\W^{5/2}}{\lambda_\text{C}^{3/2}}a_0(1+a_0^2)^{3/2}$ \\ 
\addlinespace
momentum spread & \phantom{$\displaystyle\Delta p < q$} & $\Delta p < 2\hbar k$ & $\displaystyle\frac{\Delta \gamma}{\gamma} < \frac{\rho}{\bar{\rho}}$&$\displaystyle \frac{\Delta \gamma_0}{\gamma_0}< \frac{4\gamma_0}{1+a_0^2}\frac{\lambda_\text{C}}{\lambda_\W}$ \\
\addlinespace 
quantum gain length & $L_g$ & $\displaystyle \frac{c}{2g\sqrt{N}}$ 
& $\displaystyle \frac{1}{\sqrt{\bar{\rho}}}\frac{\lambda_\W}{4\pi\rho}$ &$\displaystyle \frac{\gamma_0^2}{\sqrt{\pi}a_0\sqrt{1+a_0^2}}\sqrt{\frac{\lambda_\text{C}/
\lambda_\W}{r_\text{e} n_\text{e}}}$ \\
\addlinespace
classical gain length & $L_g^\text{cl}$ & $\displaystyle \frac{c}{\sqrt{3}} \left[\frac{2}{g^2 \omega_\text{r}N}\right]^{1/3}$ 
& $\displaystyle \frac{1}{\sqrt{3}}\frac{\lambda_\W}{4\pi\rho} $& $\displaystyle \frac{1}{\sqrt{3}}\frac{\gamma_0 \lambda_\W^{1/3}}{2\pi^{2/3}(a_0^2r_\text{e}n_\text{e})^{1/3}}$
\\
\addlinespace
plasma wave number & $k_\text{p}$  & $\displaystyle\sqrt{\frac{e^2n_\text{e}}{\varepsilon_0mc^2}}$  
& $\displaystyle\frac{4k_\W}{a_0}\rho^{3/2}$ & $\displaystyle\sqrt{\frac{4\pi r_\text{e}n_\text{e}}{\gamma_0^3}}$
\\
\addlinespace
spontaneous decay rate & $R_\text{sp}$ & $\displaystyle\frac{2\pi}{3} \frac{r_\text{e}}{\lambda_\text{C}}a_0^2 k$ & & 
 $\displaystyle\frac{(2\pi)^2}{3}\frac{a_0^2 r_\text{e}}{\lambda_\W\lambda_\text{C}}$
\\
 \bottomrule
\end{tabular}
\label{tab:bambi_lab}
\end{table*}

In Table~\ref{tab:bambi_lab} we have summarized the important parameters of our model of a Quantum FEL in the Bambini--Renieri frame as well as in the laboratory frame~\cite{peter}. 
In addition, we have expressed these quantities in terms of the `universal scaling' of Ref.~\cite{boni06}, where 
$\bar{\rho}\equiv \rho mc/q$
with  the Pierce parameter $\rho$~\cite{schmueser}, 
is the analogue of our quantum parameter $\alpha_N$~\footnote{
The universal scaling is motivated by a classical theory rather than a quantum mechanical one. 
In contrast, our approach adopts a different perspective by introducing the quantum parameter $\alpha_N$ as the ratio of the two important frequency scales of the FEL interaction. 
Moreover, $\alpha_N$ occurs in a natural way as the expansion parameter of the asymptotic expansion in Ref.~\cite{PRA2019}. However, from an experimental point of view it is often advantageous to employ the scaling of Ref.~\cite{boni06}.
}.          

\subsection{Need for optical undulator}

From Table~\ref{tab:bambi_lab} we deduce the condition 
\ba\label{eq:qhigh_alpha_lab} 
1\gg \alpha_{N} \propto a_0 (1+a_0^2)^{3/2}\,\frac{\lambda_\W^{5/2} n_\text{e}^{1/2}}{\gamma_0^3}
\ea
for the quantum regime in the laboratory frame, where $n_\text{e}$ denotes the electron density while $a_0$,  $\lambda_\W$, and $\gamma_0$ are the wiggler parameter, the wavelength of the wiggler in the laboratory frame, and  the ratio of the kinetic energy to the rest energy of an electron, respectively.

This relation implies that either increasing the electron energy $\gamma_0$, or lowering the wiggler wavelength $\lambda_\W$ leads to a decreasing value of the quantum parameter.

At the same time, we have to ensure that the gain length, 
 \ba\label{eq:qhigh_Lgain_lab} 
 L_g\propto \frac{1}{a_0(1+a_0^2)^{1/2}}\frac{\gamma_0^2}{\lambda_\W^{1/2} n_\text{e}^{1/2}}
 \ea
given in Table~\ref{tab:bambi_lab} does not become unfeasibly large.

We observe that high $\gamma_0$ as well as small $\lambda_\W$ lead to a large gain length.
However, while the scaling of $L_g$ with the energy $\gamma_0$ is quadratic, the dependence on the undulator wavelength scales only with $\lambda_\W^{-1/2}$.
Hence, in order to satisfy Eq.~\eqref{eq:qhigh_alpha_lab} and at the same time minimizing the gain length $L_g$, Eq.~\eqref{eq:qhigh_Lgain_lab}, we propose to operate a Quantum FEL with a small undulator wavelength and with a moderately high electron energy~\footnote{We note that for the classical case with  
 $L_g^{\text{cl}}\propto\gamma_0\lambda_{\text{W}}^{1/3}/(n_e a_0^{2})^{1/3}$        
from Table~\ref{tab:bambi_lab} 
we do not face the problem of an increasing gain length since the scaling with $\gamma_0$ and $\lambda_{\text{W}}$ is more favorable 
and, moreover, the condition in Eq.~(\ref{eq:qhigh_alpha_lab}) does not have to be satisfied.
}.

The requirement of a small undulator wavelength quite naturally forces us to employ an optical undulator~\cite{boni05}, where the periodic array of magnets is replaced by counterpropagating laser fields~\cite{boni05,schlicher,*schlicher_ieee,sprangle,steiniger}. 
For an efficient interaction between electrons and wiggler we also need according to Eq.~\eqref{eq:qhigh_Lgain_lab} a relatively large value of the wiggler parameter $a_0$, that is a high intensity of the counterpropagating `pump' laser.

However, a laser with the required intensity would not operate in a continuous way, but  rather in a pulsed mode which decreases the interaction length. To overcome this problem the authors in Ref.~\cite{steiniger} proposed a `traveling-wave Thompson scattering' (TWTS) scheme.
Instead of the usual head-on geometry for electrons and optical undulator, TWTS uses a side-scattering geometry, where a laser pulse with a tilted front interacts under an optimal angle with the electron beam.
This procedure can considerably enhance the interaction length.
 
\subsection{Electron beam}

Apart from the condition in Eq.~\eqref{eq:qhigh_alpha_lab} on the quantum parameter $\alpha_N$ which has led to constraints for the wiggler, we have the inequality $\Delta p < q$ for the momentum spread of the electrons which reflects itself in the required quality of the electron beam. 
According to Table~\ref{tab:bambi_lab} this condition translates into the inequality        
\begin{equation}\label{eq:qhigh_delta_gamma} 
\frac{\Delta\gamma_0}{\gamma_0}< \frac{4\gamma_0}{1+a_0^2}\frac{\lambda_\text{C}}{\lambda_\W}
\end{equation}
for the relative energy spread of the electron beam in the laboratory frame with $\lambda_\text{C}$ denoting the Compton wavelength of the electron.

Nevertheless, we note that also here a short wiggler wavelength is advantageous since it raises the upper bound of the inequality. 
Equation~\eqref{eq:qhigh_delta_gamma} represents an ambitious requirement on the quality of the electron beam~\cite{NJP2015,pio}.

Indeed, for an efficient operation of a Quantum FEL the momentum spread  $\Delta p$ has to be smaller than the  gain bandwidth $2\alpha_N q$, that is 
\begin{equation}
\label{eq:Delta_p}
\Delta p < 2\alpha_N q\,,
\end{equation} 
due to momentum selectivity~\cite{PRA2019}.
Since $\alpha_N \ll1$, Eq.~\eqref{eq:Delta_p} even lowers the maximally allowed energy spread.

Further experimental challenges due to the interaction geometry and intensity fluctuations of the pump laser as well as the required properties of the electron beam taking into account three-dimensional effects are discussed in more detail in Ref.~\cite{debus}.

\subsection{Space charge vs spontaneous emission}

So far, we have considered the  dynamics of a  Quantum FEL governed by a unitary time evolution neglecting processes which destroy the strong correlation between momentum jumps of the electrons and the emission or absorption of photons.
Decoherence mechanisms of this kind can be for example spontaneous emission into all modes of the radiation field~\cite{robb2011,*robb2012}, or space-charge effects~\cite{loui78,sprangle1} due to the Coulomb interaction between the electrons.
As these processes may eventually prevent the operation of a Quantum FEL, strong limits were imposed in Ref.~\cite{debus} on the parameter regime for the electron beam and the wiggler field.
We now show with the help of our results of Sec.~\ref{sec:Long_time_solution} that these constraints can be weakened.

The discussion in Ref.~\cite{debus} relied on estimating the typical length scales of the decoherence processes by classical theories.
Hence, we consider the rates $k_\text{p}$ and  $R_\text{sp}$, that is the plasma wave number~\cite{sprangle1,schmueser} of the electron beam and the inverse decay length~\cite{jackson,robb2011,*robb2012,rainer}, respectively, which both are listed in Table~\ref{tab:bambi_lab}.

To ensure a coherent time evolution over the total length $L$ of the wiggler, both processes have to occur on longer length scales than the interaction.
Hence, in accordance with Ref.~\cite{debus} we require the inequalities $1/(2R_\text{sp})>L$ and $1/k_\text{p}>L$, that is $R_\text{sp}L< 1/2$ and $k_\text{p}L< 1$.
In addition, we demand in the quantum regime that multi-photon processes are suppressed, that is $\alpha_N\ll 1$.

However, the parameters $k_\text{p}L$, $R_\text{sp}L$, and $\alpha_N$ are not independent of each other due to their mutual dependence on $\gamma_0$, $n_\text{e}$, $\lambda_\W$, and $a_0$.
Indeed, we can relate~\cite{debus} the wiggler length $L$   
\begin{equation}
    \label{eq:L_scsp}
\frac{L}{L_g}=\left[12 \frac{\alpha_N}{\alpha_\text{f}}(R_\text{sp}L)(k_\text{p}L)^2\right]^{1/3} 
\end{equation}
to these three parameters and the gain length.
Here we have used the definitions from Table~\ref{tab:bambi_lab} and introduced~\footnote{
For this calculation it is helpful to use the relation $g\sqrt{N}=a_0\pi c\sqrt{2r_\text{e}n_\text{e}/(\lambda_\text{C}k)}$ for the coupling constant in the Bambini--Renieri frame which can be obtained from the definitions in Ref.~\citep{NJP2015}
}
the fine-structure constant $\alpha_\text{f}\equiv 2\pi r_\text{e}/\lambda_\text{C}$.

We emphasize that the values of the parameters on the right-hand side of Eq.~\eqref{eq:L_scsp} can be chosen independently from each other.
However, once this choice is made, the interaction length on the left-hand side is fixed.  

In order to get an estimate for $L/L_g$ we set $k_\text{p}L$ and $R_\text{sp}L$ to their upper bounds, that is $k_\text{p}L=1$ and $R_\text{sp}L=0.5$, respectively.
For the example of $\alpha_N=0.25$ we obtain from Eq.~\eqref{eq:L_scsp} the value \begin{equation}
  \frac{L}{L_\text{g}}  \cong 5.9~\quad\text{(estimated limit)}
\end{equation}
for the maximally allowed interaction length.

However, from Eq.~\eqref{eq:length_seeded} we derive the saturation length \begin{equation}
  \frac{L_\text{max}}{L_\text{g}}  \cong 23.5~\quad\text{(start-up from vacuum)}
\end{equation}
for start-up from vacuum with $n_0=0$ and for $N=10^9$ electrons.
This choice for $N$ is a typical number~\cite{schmueser,pellegrini17} for electron bunches in an FEL.

Since $L \ll L_\text{max}$ the coherent time evolution breaks down long before saturation is reached, and the maximally possible intensity is extremely decreased. 

In contrast, for a seeded FEL with $n_0=0.1\,N$ we deduce from Eq.~\eqref{eq:length_seeded} the saturation length \begin{equation}
  \frac{L_\text{max}}{L_\text{g}}  \cong 5.1~\quad\text{(seeded FEL)}
\end{equation}
which is of the same order of magnitude as the allowed interaction length from Eq.~\eqref{eq:L_scsp} and thus the maximum intensity can be reached.

As a consequence, we believe in accordance with Ref.~\cite{debus} that the focus of the research on Quantum FELs should shift from SASE to seeded FELs.
We emphasize, however, that this discussion is based on arguments borrowed from classical theories and that a rigorous quantum theory covering the full interplay between multi-photon processes, spontaneous emission, and space charge is necessary to prove our statements.    

 \section{Conclusions}
\label{sec:Conclusions}

In the present article we have studied the mean intensity as well as the intensity fluctuations in a high-gain Quantum FEL. 
The reduction to two momentum levels limits the maximum intensity to a single emitted photon per electron. 
We have found that the necessary wiggler length for this maximum is significantly decreased if we consider a seeded FEL instead of SASE. 
Hence, the experimental realization of the former seems more feasible, especially with regard to the problematic requirements pointed out in Ref.~\cite{debus}. 
Our results have also shown why the short wavelength of an optical undulator is necessary 
to fulfill the most important constraints for the operation of a high-gain Quantum FEL.    

Moreover, we have observed that the time-evolved intensity fluctuations are mainly determined by the initial field state -- ranging from super- to sub-Poissonian statistics in case of a Fock or a coherent state  with a high photon number, to a very broad photon distribution for vacuum.          

To refine our model to more realistic scenarios we have to take space charge and spontaneous emission into all modes into account.
In addition, relativistic effects such as slippage~\cite{boni_sase} of the radiation pulse over the electron bunch have to be included. However, these topics go beyond the scope of our article and are subject to further studies.
 
\begin{acknowledgments}
We especially thank A.~Debus and K. Steiniger for the fruitful collaboration with emphasis on experimental aspects of FEL physics.
Moreover, we thank P. Anisimov, W. Becker, R. Endrich, A. Gover, S. Laibacher, Y. Pan, P. Preiss, 
and S. Varr{\'o} for many exciting discussions.  
W.\,P.\,S. is grateful to Texas A\&M University for a Faculty Fellowship at the 
Hagler Institute of Advanced Study at Texas A\&M University, and Texas A\&M AgriLife 
Research for the support of his work. The Research of $\mathrm{IQ}^{\mathrm{ST}}$ is 
financially supported by the Ministry of Science, Research and Art Baden-Württemberg.
\end{acknowledgments}

\begin{appendix}
\section{Analytical approximation}
  \label{sec:Analytic_approximation}

In this appendix we summarize the most important steps to obtain an analytical approximation for the mean photon number  in terms of Jacobi elliptic functions. Our procedure for Heisenberg operators is closely related to the one in Ref.~\cite{kumar} in Schwinger representation~\cite{schwinger} while it also leads to analogous results as other models~\cite{boniprep69,*boniprep70,gambini}.

\subsection{Dynamics of number operator}

Our analytical approach is based on the decoupling of the Heisenberg equations of motion for the photon-number operator $\hat{n}$ with the help of three constants of motion, in analogy to Ref.~\cite{kumar}.

The time evolution of any operator $\hat{\mathcal{O}}=\hat{\mathcal{O}}(\taub)$
in the Heisenberg picture is dictated by the Heisenberg equation of motion
\begin{equation}\label{eq:app_heisenberg}
\frac{\D}{\D\taub} \hat{\mathcal{O}}=\I\left[\hat{H}_\text{eff},\hat{\mathcal{O}}\right]   
\end{equation}
with the Hamiltonian
\begin{align}\label{eq:app_Hdicke_ang}
\hat{H}_\text{eff}\equiv\varepsilon\left(\hat{a}_\L \hat{J}_+ +\hat{a}_\L^\dagger \hat{J}_-\right)-\del \, \hat{n}\,,
\end{align}
from Eq.~\eqref{eq:Hdicke_ang}, which is independent of the time variable $\taub$.

With the help of Eqs.~\eqref{eq:app_heisenberg} and \eqref{eq:app_Hdicke_ang}, together with the commutation relations
$[\hat{J}_+,\hat{J}_-]=2\hat{J}_z$, $[\hat{J}_z,\hat{J}_\pm]=\pm\hat{J}_\pm$, and $[\hat{a}_\L,\hat{a}_\L^\dagger]=1$ we verify  that the total angular momentum 
\begin{equation}\label{eq:app_qhigh_A} 
\hat{A} \equiv\hat{\boldsymbol{J}}^2=\frac{1}{2}\left(\hat{J}_{+}\hat{J}_{-}+\hat{J}_{-}\hat{J}_{+}\right) +\hat{J}_z^2\,,
\end{equation}
the total number of excitations
\begin{equation}\label{eq:app_qhigh_B}
\hat{B} \equiv \hat{n} +\hat{J}_z\,,
\end{equation}
and the Hamiltonian
\begin{equation}\label{eq:app_qhigh_C}
\hat{C}\equiv\hat{H}_\text{eff} 
\end{equation}
are constants of motion. We note that the three constants $\hat{A}$, $\hat{B}$, and $\hat{C}$, as well as the photon-number operator $\hat{n}\equiv \hat{a}_\L^\dagger \hat{a}_\L$ mutually commute with each other.

The second derivative of $\hat{n}$ with respect to time reads
\begin{equation}
\frac{\D^2\hat{n}}{\D\taub^2} = \I \left[\hat{H}_\text{eff},\frac{\D\hat{n}}{\D\taub}\right]
=-\left[\hat{H}_\text{eff},\left[\hat{H}_\text{eff},\hat{n}\right]\right]
\end{equation}
and after inserting $\hat{H}_\text{eff}$ from Eq.~\eqref{eq:app_Hdicke_ang} and calculating the nested commutator we arrive at
\begin{equation}
 \frac{\D^2\hat{n}}{\D\taub^2} =\varepsilon^2 \left[
 \left(2\hat{n}+1\right)\hat{J}_z+ 
 \hat{J}_+\hat{J}_- +\hat{J}_+\hat{J}_-
 -\frac{\del}{\varepsilon}\left(\hat{a}_\L \hat{J}_+ +\hat{a}_\L^\dagger \hat{J}_-\right)\right]\,.   
\end{equation}
When we express the right-hand side purely in terms of the operators $\hat{A}$, $\hat{B}$, $\hat{C}$, and $\hat{n}$ we finally obtain the second-order differential equation~\cite{kumar} 
\begin{equation}\label{eq:qhigh_ddotn_op} 
\frac{\D^2\hat{n}}{\D\taub^2}=\!-2\varepsilon^2\!
\left[3\hat{n}^2\!-\!2\!\left(\!2\hat{B}-\!\left(\!\frac{\del}{2\varepsilon}\!\right)^2 \!- \! \frac{1}{2}\right)\hat{n}-\!\left(\!\hat{A}\!+\!\hat{B}\!-\!\hat{B}^2\!-\!\frac{\del\hat{C}}{2\varepsilon^2}\right)\right]
\end{equation}
for $\hat{n}$. 

The dynamics of $\hat{n}$ is indeed decoupled from the electron operators. Unfortunately, we cannot solve Eq.~\eqref{eq:qhigh_ddotn_op} by integration since it is 
a nonlinear equation of \emph{operators} instead of numbers. Hence, $\hat{n}$ and $\text{d}\hat{n}/\text{d}\taub$ do not necessarily commute.        

\subsection{Approximating operators as c-numbers}

In order to find an estimate for the mean photon number $n\equiv \braket{\hat{n}}$ we approximate the constant operators in Eq.~\eqref{eq:qhigh_ddotn_op} by their expectation values at $\taub=0$, that is $A=r(r+1)$, $B=n_0+m$, and $C=-\del n_0$ with $r=m=N/2$. Here we have assumed that the field starts with the photon number $n=n_0$, and initially all electrons are in the excited state close to $p=q/2$. 
Strictly speaking, this approximation is only valid as long as products of operators result in products of expectation values~\cite{kumar} when we form the total expectation value of Eq.~\eqref{eq:qhigh_ddotn_op}.

Then, we multiply the resulting c-number equation by $\text{d}n/\text{d}\taub$ and integrate  over time $\taub$. This procedure yields the equation of motion
\ba\label{eq:qhigh_dn2} 
\left(\frac{\D n}{\D\taub}\right)^2=4\varepsilon^2 (n_+-n)(n-n_0)(n+n_-)
\ea
where 
\begin{equation}\label{eq:qhigh_npm}
 \begin{aligned} 
n_{\pm}\equiv \pm \frac{N}{2}\!\left(1-\frac{\del^2}{4\alpha_N^2}\right)\pm \frac{1}{2}\!\left(n_0 - \frac{1}{2}\right)\!+\frac{1}{2}
\!\left[N^2\!\left(1-\frac{\del^2}{4\alpha_N^2}\right)^2\right. \\ 
\left.+\left(n_0+\frac{1}{2}\right)^2
+2N n_0\left(1+\frac{\del^2}{4\alpha_N^2}\right)+3N\left(1+\frac{\del^2}{12\alpha_N^2}\right)\right]^{1/2}
 \end{aligned}
\end{equation}
denotes the roots of the right-hand side of Eq.~\eqref{eq:qhigh_dn2}. 

By setting $\D n/\D\taub=0$ in Eq.~\eqref{eq:qhigh_dn2} we observe that the maximum photon number is given by $n_+$. The other two roots of Eq.~\eqref{eq:qhigh_dn2}, that is $n=n_0$ and $n=-n_-$, correspond to the minimum and initial value of $n$, and to an unphysical negative photon number, respectively.

We proceed by integrating Eq.~\eqref{eq:qhigh_dn2} and arrive at the expression  
\begin{equation}\label{eq:qhigh_elliptic_intsol} 
2\alpha_N \taub =\int\limits_{n_0/N}^{n/N}\!\!\frac{\D y}{\sqrt{(n_+/N-y)(y-n_0/N)(y+n_-/N)}}
\end{equation}
which describes an elliptic integral.

 \subsection{Solution of elliptic integral}
\label{app:Solution_of_elliptic_integral}

Next we invert the elliptic integral from Eq.~\eqref{eq:qhigh_elliptic_intsol}
in order to express the mean photon number $n$ as a function of the dimensionless time $\taub$.
Our result is then presented in terms of Jacobi elliptic functions~\cite{byrd}. 

These special functions are defined as inverse functions of the elliptic integral of first kind
\ba 
u(\varphi,\mathfrak{K})\equiv \int\limits_0^\varphi \frac{\text{d}y}{\sqrt{1-\mathfrak{K}^2\sin^2{y}}}
\ea
which is characterized by the amplitude $\varphi$ and the modulus $\mathfrak{K}$ with $0<\mathfrak{K}<1$. 

There exists a set of  elliptic functions, for example $\text{sn}\equiv \sin{\varphi}$ and $\text{cn}\equiv \cos{\varphi}$ which are called sine amplitude and cosine amplitude, respectively. They vary between $-1$ and $1$ and are $4K$-periodic, with  
$K(\mathfrak{K})\equiv u(\pi/2,\mathfrak{K})$ denoting a \textit{complete} elliptic integral of first kind~\cite{byrd}. 

We note that $K$ shows the asymptotic behavior~\cite{byrd} 
\ba\label{eq:app_K_asympt} 
K\cong \ln{\left(\frac{4}{\sqrt{1-\mathfrak{K}^2}}\right)}
\ea
for $\mathfrak{K}$ approaching unity, that is $\mathfrak{K}\rightarrow 1$.
 
We now return to the solution of the integral in Eq.~\eqref{eq:qhigh_elliptic_intsol}
and observe for $N\gg n_0$ and $\del/\alpha_N< 2$ the relative ordering  $n_+\geq n \geq n_0 > -n_-$ for the roots of the denominator. Following  Ref.~\cite{byrd} we proceed by performing the substitution 
\begin{align}\label{eq:app_jac_subst} 
\sn^2(u,\mathfrak{K})&=\frac{1}{\mathfrak{K}^2}\frac{y-n_0/N}{y+n_-/N}\,, 
\end{align}
where
\begin{align}\label{eq:app_jac_ksubst} 
\mathfrak{K} &\equiv \sqrt{\frac{n_+-n_0}{n_++n_-}}
\end{align} 
denotes the modulus corresponding to the $\sn$ function.

With the help of the substitution in Eqs.~\eqref{eq:app_jac_subst}  and~\eqref{eq:app_jac_ksubst} we perform the integration in Eq.~\eqref{eq:qhigh_elliptic_intsol}. Solving the resulting expression for $n$ leads us finally to
\begin{equation}\label{eq:app_nL_jac} 
n=n_0+(n_+-n_0)\,\cn^2\left(\frac{L}{2L_g}\sqrt{\frac{n_++n_-}{N}}-K,\mathfrak{K}\right)
\end{equation}
after using several fundamental identities~\cite{byrd} for Jacobi elliptic functions. 
Moreover, we have expressed our result in terms of the length $L\equiv ct$ and the gain length $L_g\equiv c/(2g\sqrt{N})$ by recalling the relation $\alpha_N \tau = L/(2L_g)$ from Ref.~\cite{PRA2019}.

  \section{Numerical solution}
 \label{sec:Numerical_solution}

In contrast to our approach in App.~\ref{sec:Analytic_approximation} which is based on time-dependent operators, we now employ for  our numerical solution time-dependent state vectors following Ref.~\cite{walls70}. We first address the case of an initial Fock state and then turn to an arbitrary initial state to obtain moments of the photon distribution of a Quantum FEL.

\subsection{Three-term recurrence relation}

The suitable basis for the electrons in this description is given by the states $\ket{r,m}$ which fulfill the eigenvalue equations~\cite{ct}
\ba\label{eq:qhigh_ang_eigen} 
\hat{\boldsymbol{J}}^2\ket{r,m}&=r(r+1)\ket{r,m} \ \ \ \text{and} \ \ \
\hat{J}_z \ket{r,m}=m\ket{r,m}\,.
\ea
Hence, the positive integer $r$ corresponds to the total angular momentum and $m$ to its $z$-component. 

Moreover, we recall~\cite{ct} the relation
\ba\label{eq:qhigh_ladder}
\hat{J}_{\pm}\ket{r,m}= \sqrt{(r \pm m +1)(r \mp m)}\ket{r,m \pm 1}
\ea
for the action of the ladder operators $\hat{J}_{\pm}$ on such a state. 

Therefore, we expand the combined state vector $\ket{\Psi}=\ket{\Psi(\tau)}$ of the laser field and the electrons into the basis $\ket{n,r,m}$. Here $\ket{n}$ denotes a Fock state of the laser field with $n$ photons while the eigenstate $\ket{r,m}$ for angular momentum describes the electrons.

Since the total angular momentum $\hat{A}$ as well as the total number of excitations $\hat{B}$, Eq.~\eqref{eq:qhigh_AB}, are constants of motion, we deduce that  $r=N/2=\text{const}$ and $m+n=N/2+n_0=\text{const}$, respectively. 

Hence, we expand the total state vector~\cite{walls70}
\begin{equation}\label{eq:qhigh_state_ang} 
\ket{\Psi(\taub)}\equiv \sum\limits_{\nu=n_0}^{n_0+N} c_\nu(\taub)\left|{\,\nu,\frac{N}{2},\frac{N}{2}+n_0-\nu\,}\right\rangle
\end{equation}
in terms of the single quantum number $\nu$  with the expansion coefficients $c_\nu=c_\nu(\tau)$. We note that $\nu$ varies between  $\nu=n_0$ and $\nu=n_0+N$ due to the constraint $-N/2 \leq m \leq N/2$.

With the help of Eq.~\eqref{eq:qhigh_ladder} and the Schr{\"o}dinger equation governed by the  Hamiltonian from Eq.~\eqref{eq:Hdicke_ang} we derive the equation of motion 
\begin{equation}\label{eq:qhigh_three_term} 
\I\frac{\D c_\nu}{\D\taub}=-\del \, \nu\, c_\nu + \alpha_N
\left[a(\nu)\,c_{\nu-1}+a(\nu+1)\,c_{\nu+1} \right]
\end{equation}
for the probability amplitudes $c_\nu$ where the coefficients
\begin{equation}
a(\nu)\equiv \sqrt{\nu(\nu-n_0)}\sqrt{1-\frac{\nu-n_0-1}{N}}
\end{equation}
are independent of $c_\nu$.

The solution of this linear system of differential equations relies on the diagonalization of a tridiagonal matrix with the dimension $(N+1)\times (N+1)$ which can be straightforwardly
achieved by numerical methods. For this purpose, we choose the initial state $\ket{n_0,N/2,N/2}$, that is all electrons initially are in the excited state while the laser field starts from $n=n_0$, which leads to the initial condition $c_\nu(0)=\updelta_{\nu,n_0}$ for Eq.~\eqref{eq:qhigh_three_term}.  

\subsection{Arbitrary initial state}

So far, we have only considered a Fock state as initial field state. We easily generalize our approach to an arbitrary initial state characterized by the photon statistics $p_n$. 

In the present article, we are mainly interested in the photon statistics  
\begin{equation}\label{eq:pn_coh}
     p_n^\text{coh}=\frac{n_0{}^n}{n!}\e{-n_0} 
\end{equation}
and
\begin{equation}\label{eq:pn_therm}
    p_n^\text{therm}=\frac{1}{n_0+1}\left(\frac{n_0}{n_0+1}\right)^n
\end{equation}
of a coherent state and a thermal state  both characterized by the mean photon number $n_0$.

The time-evolved density operator $\hat{\varrho}=\hat{\varrho}(\tau)$ of the combined system of electrons and laser field reads
\begin{equation}\label{eq:rho_t}
\hat{\varrho}(\taub)=\sum\limits_{k,l=0}^\infty \varrho_{k,l}
\sum\limits_{\mu=k}^{N+k}\sum\limits_{\nu=l}^{N+l}c_\mu(\tau)c_\nu^*(\tau)
\hat{\mathcal{M}}(\mu,k|\nu,l)
\end{equation}
with the operator
\begin{equation}
\hat{\mathcal{M}}(\mu,k|\nu,l)\equiv \ket{\mu,N/2,N/2+k-\mu}\bra{\nu,N/2,N/2+l-\nu}\,.
\end{equation}
The coefficients $\varrho_{k,l}$ denote the matrix elements of the initial density operator for the laser field in photon number representation. The diagonal elements of this matrix define the photon statistics, that is $p_n\equiv \varrho_{n,n}$.

In order to calculate the expectation value of any function $f=f(\hat{n})$ of the number operator $\hat{n}$ we trace over the electrons and the laser field, that is
\begin{equation}
\braket{f(\hat{n})}=\traceind{\text{L}}{\traceind{\text{el}}{\hat{\varrho}(\taub)}f(\hat{n})}\,,
\end{equation}
and arrive at the relation
\begin{equation}
\braket{f(\hat{n})}=\sum\limits_{n=0}^\infty \bra{n}
\left(\sum\limits_{m=-N/2}^{N/2}\bra{N/2,m}\hat{\varrho}(\taub)\ket{N/2,m}\right)f(n)\ket{n}
\end{equation}
for this expectation value.

After inserting the expression for $\hat{\varrho}$ in Eq.~\eqref{eq:rho_t}, and calculating the matrix element 
\begin{equation}
\left\langle{n,\frac{N}{2},m}\right|\hat{\mathcal{M}}(\mu,k|\nu,l)\left|{n,\frac{N}{2},m}\right\rangle =
\updelta_{n,\nu}\updelta_{m,N/2+l-\nu}\,\updelta_{\mu,\nu}\updelta_{k,l}
\end{equation}
we obtain the expression
\begin{equation}\label{eq:f_zwischen}
\braket{f(\hat{n})}=\sum\limits_{n'=0}^\infty p_{n'}\,
\sum\limits_{\nu=n'}^{N+n'}|c_\nu(\taub)|^2 f(\nu)
\end{equation}
which only contains the diagonal elements $p_{n'}$. 

When we set $p_{n'}=\updelta_{n',n}$ which corresponds to a Fock state, Eq.~\eqref{eq:f_zwischen} reduces to the formula
\begin{equation}
\braket{f(\hat{n})}_n\equiv 
\sum\limits_{\nu=n}^{N+n}|c_\nu(\taub)|^2 f(\nu)
\end{equation}
corresponding to the case with initial photon number $n$.This result also emerges by solving Eq.~\eqref{eq:qhigh_three_term}, and calculating the expectation value of $f(\hat{n})$ with respect to the state in Eq.~\eqref{eq:qhigh_state_ang}.

Hence, we finally obtain the elementary expression  
\begin{equation} \label{eq:fn_av}
\braket{f(\hat{n})}=\sum\limits_{n'=0}^\infty p_{n'}\braket{f(\hat{n})}_{n'}
\end{equation}
for the expectation value of  $f=f(\hat{n})$. 

This result arises due to our choice of the initial state $\ket{N/2,N/2}$ of the electrons and due to the fact that there is only one independent quantum  number, $\nu$. \end{appendix}

\bibliography{bib}

\end{document}